\documentclass[useAMS,usenatbib]{mn2e}
\usepackage{graphicx}

\title[The magnetic field of the planet-hosting star $\tau$~Bootis]
{The magnetic field of the planet-hosting star $\tau$~Bootis
\thanks{Based on observations 
obtained at the Canada-France-Hawaii Telescope (CFHT) which is operated by 
the National Research Council of Canada, the Institut National des Sciences 
de l'Univers of the Centre National de la Recherche Scientifique of France, 
and the University of Hawaii.}
}
\author[C. Catala et al.]{C. Catala$^{1}$ 
\thanks{E-mail:
Claude.Catala@obspm.fr}, 
J.-F. Donati$^2$\thanks{E-mail: Jean-Francois.Donati@obs-mip.fr},
E. Shkolnik$^3$\thanks{E-mail: shkolnik@ifa.hawaii.edu},
D. Bohlender$^4$\thanks{E-mail: david.bohlender@nrc-cnrc.gc.ca}, 
E. Alecian$^{1}$\thanks{E-mail: Evelyne.Alecian@obspm.fr}\\
$^1$ Observatoire de Paris, LESIA, 5 place Jules Janssen, 92195 Meudon Cedex, France \\
$^2$ Observatoire Midi-Pyr\'en\'ees, LATT, 14 avenue Edouard Belin, 31400 Toulouse, France \\
$^3$ NASA Astrobiology Institute, University of Hawaii at Manoa, 2680 Woodlawn Drive, Honolulu, HI, 96822 \\
$^4$ National Research Council of Canada, Herzberg Institute of Astrophysics, 5071 West Saanich Road, Victoria, BC V9E 2E7, Canada }

\begin{document}

\date{Accepted . Received ; in original form }

\pagerange{\pageref{firstpage}--\pageref{lastpage}} \pubyear{2006}

\maketitle

\label{firstpage}

\begin{abstract}
We have obtained high resolution spectropolarimetric data for the 
planet-hosting star $\tau$~Bootis, using the ESPaDOnS spectropolarimeter
at CFHT. A weak but clear Stokes $V$ signature is detected on three of the four
nights of June 2006 during which we have recorded data. This polarimetric
signature indicates with no ambiguity the presence of a magnetic field at
the star's surface, with intensity of just a few Gauss.

The analysis of the photospheric lines of $\tau$~Boo at ultra-high
signal-to-noise reveals the presence of an 18\% relative differential 
rotation. Tentative Zeeman-Doppler imaging, using our spectropolarimetric 
observations covering only a fraction of the star's rotational phase, 
indicates a magnetic field with a dominant potential field component.
The data are best fitted when a 3.1d period of modulation
and an intermediate inclination are assumed. Considering the level of
differential rotation of $\tau$~Boo, this implies a rotation period of 3.0d
at the equator and of 3.7d at the pole, and a topology of the magnetic field
where its main non-axisymmetric part is located at low latitudes. 

The planet is probably synchronised with the star's rotation at 
intermediate latitudes, while the non-axisymmetric part of the magnetic 
field seems located at lower latitudes. Our limited data do not provide 
sufficient constraints on the magnetic field to study a possible 
interaction of the planet with the star's magnetosphere. Investigating 
this issue will require data with much
better phase coverage. Similar studies should also be performed for 
other stars hosting close-in giant planets.

\end{abstract}


\section{Introduction}
It has been recently conjectured that giant planets in close-in orbits can 
influence significantly the magnetic activity of their parent stars, 
through tidal interaction, or via magnetic coupling between the star's and the
planet's fields (Cuntz et al. 2000, Rubenstein \& Schaefer 2000). Such a
scenario is strongly suggested in the case of HD 192263, which has a planet 
with $M \sin i = 0.72 M_J$ orbiting at 0.15 AU in 24.3d, by the cyclical 
photometric variations with a similar period, stable over almost 4 years
(Henry et al. 2002, Santos et al. 2003). In addition, Shkolnik et al. (2003,
2005)
monitored chromospheric activity of several giant planet-hosting stars in the
Ca~{\sc II} H \& K lines, and found clear evidence for cyclical variations 
of chromospheric signatures synchronised with the planet orbits in the case of
two of them, HD 179949 (planet with $M \sin i = 0.98 M_J$, $P_{orb}=3.09$d, 
semi-major axis = 0.045 AU) and $\upsilon$~And 
(planet with $M \sin i = 0.71 M_J$, 
$P_{orb}=4.62$d, semi-major axis = 0.059 AU). These authors offer an
interpretation of the interaction between the planet and the star's magnetic
field, in which Alfv\'en waves are generated by the slow relative azimuthal
motion of the planet with respect to the stellar magnetic field.

The star $\tau$~Boo (F7V), which has a giant planet 
companion orbiting in 3.31 d at 0.049 AU with a minimum mass 
$M \sin i$ of 4.4 $M_J$ (Butler et al. 1997, Leigh et al. 2003), 
does not show such synchronised
Ca~{\sc II} H \& K line variations. Shkolnik et al. (2005) argue that this is 
consistent with their Alfv\'en wave model, if the star is tidally locked
by its hot giant planet. Ca~{\sc II} H \& K spectrophotometric monitoring 
of $\tau$~Boo during several seasons indeed suggests rotation periods 
ranging between 2.6 and 4.1d, depending on the season (Henry et al. 2000). 
This large interval is indicative of a possible differential rotation 
at the star's surface, active regions being located 
at different latitudes during 
different seasons. The planet orbital period is therefore possibly
identical to the rotation period of the star at a particular latitude. 
Moreover, recent ultra-high 
precision photometric monitoring using the MOST asteroseismology satellite 
(Walker et al. 2003, Matthews et al. 2004) indicates the possible 
presence of an active spot located 
near the sub-planetary longitude, stable over more than 100 orbits 
(Walker et al. 2005). Finally, high precision linear polarisation monitoring
of $\tau$~Boo reveals very low amplitude modulation of the fractional 
linear polarisation, although the origin of such modulation is still unclear
(Lucas et al. 2006).

Clearly, the direct detection of a magnetic field at the surface of
$\tau$~Boo, and the determination of its topology would constitute 
a powerful way of investigating the possible relationship between the
planet and the star's magnetism. This would be a first step of a more complete
statistical study of magnetic fields in stars hosting hot giant exoplanets, 
aimed at characterizing the interaction of exoplanets with the magnetic field 
of their parent stars.

This paper presents the first direct detection of the magnetic field of
$\tau$~Boo, and a tentative mapping of its topology from our fragmentary
data set. Section 2 describes the 
observations and data reduction. In Sect. 3, we present the results of our
analysis, concerning the differential rotation of the star's surface, the
magnetic field intensity and topology, as well as the rotation period
and inclination angle of the star. Section 4 
provides a general conclusion and comments on future prospects
of this work.

\section{Observations and data reduction}
We used the ESPaDOnS spectropolarimeter installed on the 3.6m 
Canada-France-Hawaii Telescope (Donati et al. 2006, in preparation). The star 
$\tau$~Boo was observed during 4 nights in June 2006, each night for durations
ranging from 20 to 60 minutes. Table 1 presents the log of the observations.

\begin{table}
 \centering
  \caption{Journal of ESPaDOnS observations of $\tau$~Boo. The last 
column gives the peak signal-to-noise ratio per velocity bin of 
2.6~km\thinspace s$^{-1}$, obtained at wavelengths near 700 nm.
The orbital phases are calculated using the ephemeris of Collier Cameron 
(private communication, see text). The phase origin is taken as the time of
planet opposition.}
  \begin{tabular}{|l|l|l|l|l|}
  \hline
   UT date     &  HJD        & orbital &  t$_{\rm exp}$   &   S/N     \\
 (dd/mm/yy)    & (2400000+)  & phase   &      (s)         &           \\
\hline
13/06/06       & 53899.75127 & 0.4785  &   800            &   2000    \\
13/06/06       & 53899.76233 & 0.4820  &   600            &   1700    \\
13/06/06       & 53899.77276 & 0.4855  &   600            &   1700    \\
13/06/06       & 53899.78216 & 0.4879  &   600            &   1700    \\
13/06/06       & 53899.79145 & 0.4914  &   600            &   1700    \\
13/06/06       & 53899.80073 & 0.4938  &   600            &   1700    \\
17/06/06       & 53903.78645 & 0.6966  &   600            &   1700    \\
17/06/06       & 53903.79580 & 0.7002  &   600            &   1700    \\
18/06/06       & 53904.78201 & 0.9974  &   600            &   1600    \\
18/06/06       & 53904.79143 & 0.0009  &   600            &   1600    \\
19/06/06       & 53905.83950 & 0.3169  &   1200           &   1500    \\
19/06/06       & 53905.85570 & 0.3217  &   1200           &   1400    \\
\hline
\end{tabular}
\end{table}

The data were obtained in the polarimetric configuration of ESPaDOnS,
yielding a spectral resolution of about 65,000. All spectra were recorded as
sequences of 4 individual subexposures taken in different configurations of
the polarimeter, in order to yield a full circular polarisation analysis,
as described in Donati et al. (1997) and Donati et al. (2006, in preparation).
No linear polarisation analysis was performed. The data were reduced with
the automatic reduction package "Libre-ESpRIT" installed at CFHT (Donati et
al.  1997, Donati et al. 2006 in preparation). Stokes $I$ and Stokes $V$
spectra are obtained by proper combinations of the 4 subexposures, while
check spectra, labelled as $N$ spectra, are calculated by combining
the subexposures in such a way to yield a null spectrum, that can be
used to verify the significance of the signal measured in Stokes $V$.

We subsequently applied the Least-Square Deconvolution (LSD) method described
in Donati et al. (1997) to construct average photospheric profiles both 
of the Stokes $I$ and $V$ parameters. The LSD technique builds the mean 
photospheric line profile, both in Stokes $I$ and $V$
by deconvolving the observed spectrum from a line mask including all lines 
present in a synthetic spectrum of the star. The line mask used here
was computed using a Kurucz Atlas 9 model with T$_{\rm eff} = 6250$~K 
and Log $g$ = 4.0, and includes about 4,000 lines with depth relative to 
the continuum larger than 0.4.

We measured the heliocentric radial velocity of $\tau$~Boo by fitting a 
gaussian to the LSD $I$ mean profile. The star's reflex motion due to the 
planet revolution is clearly seen in our V$_{\rm rad}$ measurements,
which agree with the planet ephemeris: 
JD$_0$ = 2453450.984 (time of opposition); P$_{\rm orb}$ = 3.31245 d;
amplitude = 0.474 km \thinspace s$^{-1}$, eccentricity = 0, provided by 
Collier Cameron (private communication), with a 
residual dispersion of about 20 m \thinspace s$^{-1}$. This ephemeris is
within the error bar of that published by Shkolnik et al. (2005). 

All Stokes $I$ and $V$ LSD profiles were subsequently converted to the
star's rest frame by correcting the velocity scale for the 
orbital motion.

\section{Results}

\subsection{Fundamental parameters of $\tau$~Boo}
Most basic fundamental parameters of $\tau$~Boo are compiled in Table 1 of 
Leigh et al. (2003), and we adopt them in the present paper: spectral type F7V; 
T$_{\rm eff} = 6360 \pm 80$~K; $[Fe/H] = 0.27 \pm 0.08$; 
$M_V = 4.496 \pm 0.008$; mass $M_* = 1.42
\pm 0.05$~M$_{\odot}$; radius $R_* = 1.48 \pm 0.05$~R$_{\odot}$; age
$1.0 \pm 0.6$~Gyr.

The rotational parameters of $\tau$~Boo are less well known, in particular
its rotation period $P_{rot}$, which is believed to be between 2.6 and 4.1d 
(Henry et al. 2000). Several authors have suggested that the star's rotation
and the planet orbital motion are tidally locked, implying a rotation period
of 3.31d for the star (Leigh et al. 2003, Collier Cameron \& Leigh 2004, 
Shkolnik et al. 2005). This hypothesis has never been directly verified 
observationally. The photospheric line profiles are reasonably well 
fitted assuming a turbulent velocity of 5.5 km\thinspace s$^{-1}$ and an 
homogeneous surface rotation with 
$v \sin i = 14.5 \pm 0.1$ km\thinspace s$^{-1}$ (see Sect. 3.2). 

These values and error bars for $P_{rot}$, $v \sin i$ and $R_*$ indicate an 
intermediate inclination of the rotation axis with respect to the line 
of sight, $30^{\circ} \leq i \leq 60^{\circ}$, the main uncertainty being 
related to the large error bar on $P_{rot}$. An inclination angle 
$i = 40^{\circ}$, corresponding to a rotation period identical to the
planet orbital period, is often considered as the most probable value 
(Leigh et al. 2003, Collier Cameron \& Leigh 2004). However, a better 
direct determination of $P_{rot}$ would be of great importance, as it 
would allow us to measure the inclination angle $i$ with a better accuracy 
and reliability. 
Since the star's rotation axis and the planet orbital axis are certainly
aligned, this would increase the accuracy of the planet mass determination. 

\subsection{The differential rotation of $\tau$~Boo}

\begin{figure}
\centering
\includegraphics[width=7cm,angle=-90]{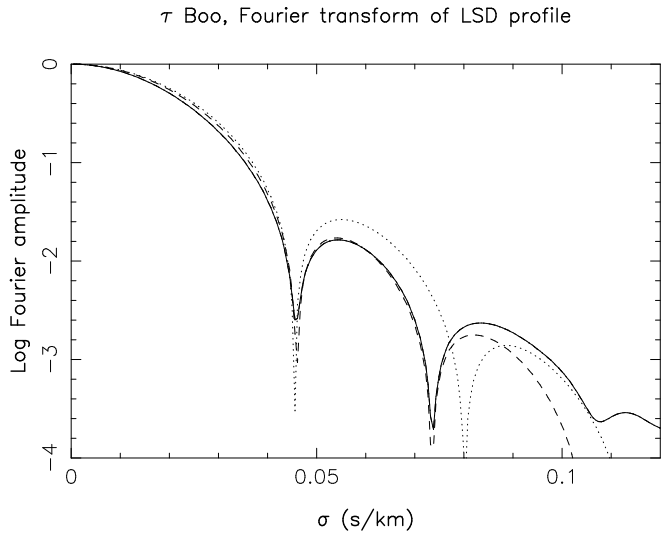}
\caption[]{Fourier transform of the LSD average photospheric profile
of $\tau$~Boo (solid line), compared to theoretical profiles assuming a
solid rotation with $v \sin i$ = 14.5 km\thinspace s$^{-1}$ (dotted line)
and a differential rotation with a projected equatorial rotation velocity 
$V_{eq} \sin i = 15.9$~km\thinspace s$^{-1}$, a projected polar 
rotation velocity $V_{pole} \sin i  = 13.0$~km\thinspace s$^{-1}$, and 
an inclination angle $i = 40^{\circ}$ (dashed line).}
\label{fft}
\end{figure}

We averaged all Stokes $I$ profiles obtained during this run, evaluated
in the star's rest frame,
and calculated the Fourier transform of the resulting very high signal-to-noise 
profile, in order to analyze the projected rotation and potentially the 
surface differential rotation of $\tau$~Boo, using the method described in 
Reiners \& Schmitt (2003). These results are presented in Fig.~\ref{fft},
where the Fourier transform of the observed profile is compared to both
rigidly rotating and differentially rotating models. 
We see clearly the first
two zeros of the Fourier transform, and measure a ratio $q_2/q_1 = 1.60 \pm
0.02$ of the positions of these zeros. This value agrees well with, but
has a better precision than that 
measured by Reiners (2006), $q_2/q_1 = 1.57 \pm 0.04$.
Rigidly rotating models
include a turbulent velocity of 7.0 km\thinspace s$^{-1}$ (FWHM = 11.7
km\thinspace s$^{-1}$), necessary to
reproduce the photospheric profile, while the needed turbulent velocity
in the case of differential rotation is only 5.5 km\thinspace s$^{-1}$ 
(FWHM = 9.2 km\thinspace s$^{-1}$). 

This value of $q_2/q_1$, as well as the direct comparison of the Fourier 
transform of the observed profile with theoretical profiles, clearly 
demonstrate the presence of a significant surface differential rotation. 
The amount of differential rotation necessary to reproduce the photospheric
profile depends on the inclination angle, which is not well known. With
an inclination $i = 40^{\circ}$, which is the most probable value
obtained when the star's rotation and planet orbital motion
are assumed tidally locked, the photospheric profile can be reproduced 
assuming an 18\% relative differential rotation, with an equator (resp. a pole)
projected rotation velocity of 
15.9 km\thinspace s$^{-1}$ (resp. 13.0 km\thinspace s$^{-1}$). This
corresponds to a differential rotation $d\Omega \approx 0.4$~rad.d$^{-1}$. A
lower inclination $i = 30^{\circ}$, which is not ruled out for $\tau$~Boo 
(see Sect. 3.1), would lead to a relative differential 
rotation of only 15\%, with $V_{eq} \sin i = 15.9$ km\thinspace s$^{-1}$ and 
$V_{pole} \sin i = 13.5$ km\thinspace s$^{-1}$, i.e. a differential rotation
$d\Omega \approx 0.3$~rad.d$^{-1}$. 

The photospheric spots accompanying the magnetic structure described 
below may produce an asymmetric photospheric profile, which could impact the
determination of differential rotation, as noted by Reiners \& Schmitt (2003). 
However, we have verified that there is very little asymmetry in the case 
of $\tau$~Boo, by comparing the profile to a mirror version of itself. Besides,
the very similar values for the ratio $q_2/q_1$ found by Reiners (2006) and
ourselves at randomly selected epochs provides further evidence that spots
induce very little variability on the photospheric profile. We conclude that
our determination of differential rotation in $\tau$~Boo is not significantly
affected by photospheric spots. 

\subsection{Magnetic field}

\begin{figure}
\centering
\includegraphics[width=7cm,angle=-90]{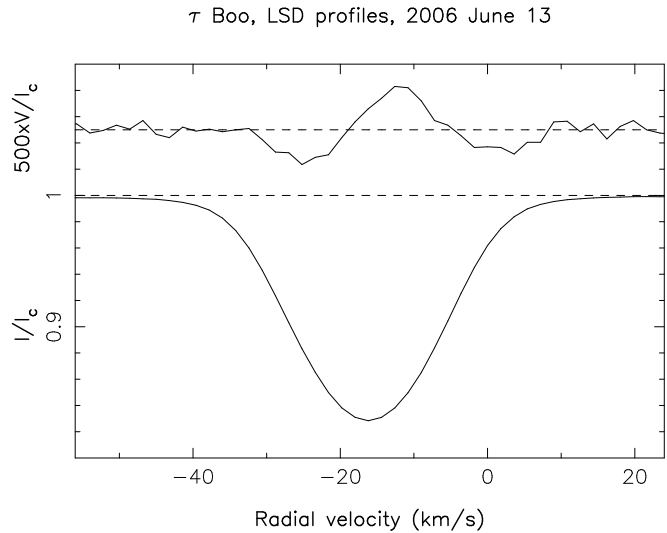}
\caption[]{LSD unpolarised and circularly polarised profile of $\tau$~Boo,
recorded on June 13, 2006. Note that the Stokes $V$ profile is shifted
vertically and expanded by a factor 500 for display purposes.}
\label{lsd}
\end{figure}

We detect a clear Stokes $V$ signature on June 13, 17 and 
19, 2006, while no Stokes $V$ signal is seen in the spectra of 
June 18, 2006. The noise level in the LSD Stokes $V$ profiles is of the order 
of 2 to 3 $\times 10^{-5}$ per spectrum, with a multiplex gain of about 25 in 
signal-to-noise ratio from the simultaneous use of all lines in the LSD mask.
The equivalent noise level taking all recorded spectra into account is lower 
than $10^{-5}$. As an example, Fig.~\ref{lsd} shows one of the Stokes $I$ 
and $V$ LSD profiles recorded on June 13, 2006, in which the Stokes $V$ 
signal is clearly detected at more than 10$\sigma$, implying with no ambiguity 
the presence of a surface magnetic field. 

Although our observations cover only a small fraction of the rotational phase
of $\tau$~Boo, we attempted to reconstruct this magnetic field from the set 
of Stokes $V$ profiles at our disposal. We used the method fully described 
in Donati et al.
(2006), based on maximum-entropy image reconstruction of the field topology,
approximated by a sum of successive spherical harmonics. We chose to limit
the spherical harmonic decomposition to $\ell_{max} = 10$, after checking that 
the results are basically unchanged for all values of $\ell_{max} \geq 5$.
Note that the projected rotation velocity at the equator of $\tau$~Boo 
($V_{eq} \sin i$ = 15.9 km\thinspace s$^{-1}$) is large enough to obtain more
than 10 resolved elements around the star at the equator, so that the
reconstruction should be accurate for spherical harmonics with 
$\ell \leq 5$, whereas the power of the reconstructed image on higher order 
harmonics is underestimated.

Because the inclination angle $i$ and the star's rotation period $P_{rot}$ 
are not well constrained, we reconstructed magnetic images for several values
of both parameters, in the range 30--60$^{\circ}$ for $i$ and 2.6--4.1d for
$P_{rot}$. The interval for $P_{rot}$ is that given by
Henry et al. (2000) for the probable photometric period of the star, while
that for $i$ is discussed in Sect. 3.1. For each test couple ($i$, $P_{rot}$), 
we reconstructed the maximum-entropy magnetic image as in Donati et al. (2006), 
compared the set of calculated Stokes $V$ profiles to the observed ones, 
and calculated the corresponding reduced $\chi^2_{\nu}$. We first explored 
solutions involving only potential fields, then introduced additional 
toroidal components in a second step. For the calculation of
the reconstructed profiles, we used a gaussian macroturbulence of 7 
km\thinspace s$^{-1}$, which leads to an excellent fit to the Stokes $I$ 
profile when no differential rotation is introduced, as seen in Sect. 3.2.

\begin{figure}
\centering
\includegraphics[width=6cm]{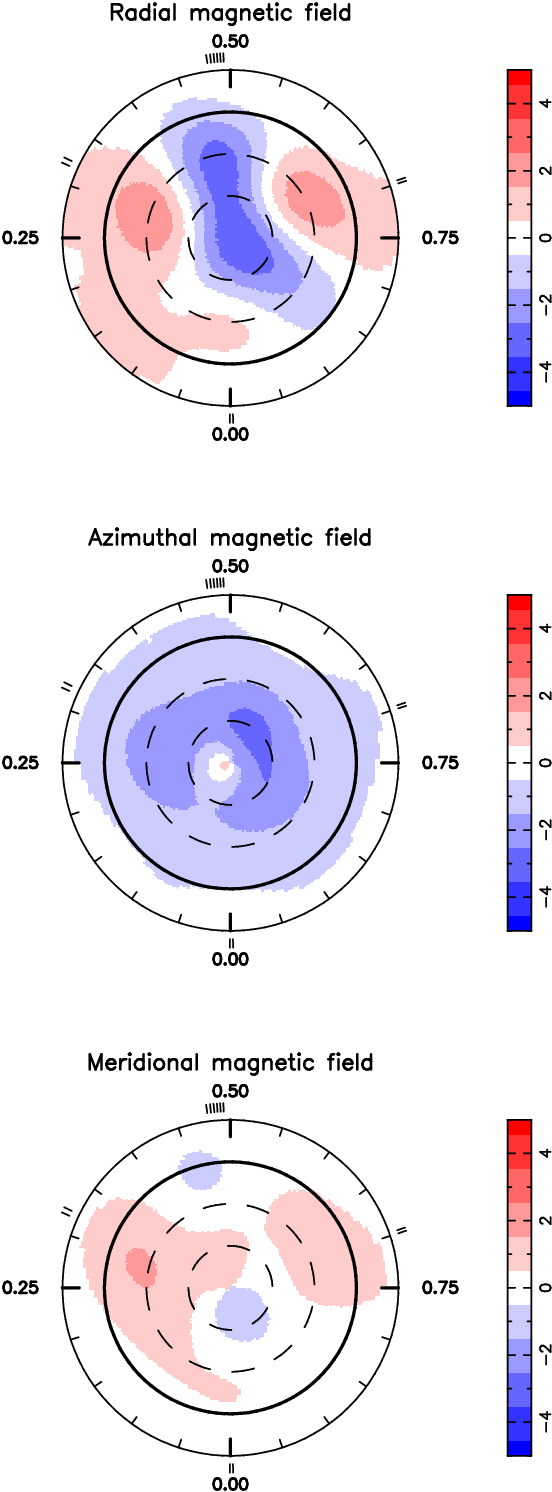}
\caption[]{
Maximum-entropy reconstructions of the magnetic
topology of $\tau$~Boo, assuming that the global field can be expressed
as the sum of a potential field and a toroidal field. The three components
of the field are displayed from top to bottom (flux values
labelled in G); the star is shown in flattened polar projection
down to latitudes of -30$^{\circ}$ with the equator depicted as a bold
circle and parallels as dashed circles. Radial ticks around each
plot indicate orbital phases of observations, counted from the time of planet 
opposition.}
\label{map}
\end{figure}

\begin{figure}
\centering
\includegraphics[width=10cm,angle=-90]{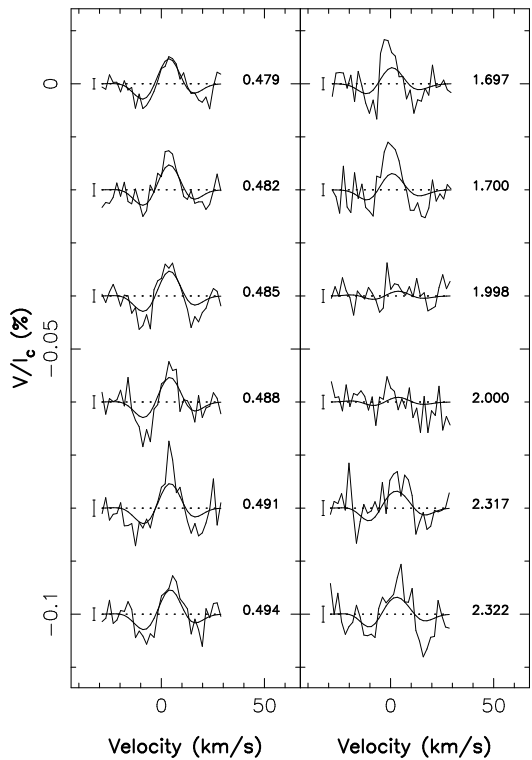}
\caption[]{Maximum-entropy fit (thick lines) to the observed Zeeman signatures 
of $\tau$~Boo (thin lines). The planet orbital phase and cycle of
each observation are written on the right side of each profile. The phase
origin is the time of planet opposition. 1$\sigma$ 
error bars are shown on the left side.}
\label{fit}
\end{figure}

When no differential rotation is included, we find a clear minimum of 
$\chi^2_{\nu}$ at a period of $3.1 \pm 0.1$d. Figure~\ref{map} shows the 
reconstructed magnetic image of $\tau$~Boo, 
assuming this rotation period and an inclination angle 
$i = 40^{\circ}$, while Fig.~\ref{fit} displays the corresponding fit to the 
data. Reconstructed maps for the same value of $P_{rot}$ and other values 
of the inclination between 30 and 60$^{\circ}$ are only slightly different.

We find that the model represented by the reconstructed image leads to
a fit of the circular polarisation data with a reduced $\chi^2_{\nu}$ of 1, 
while the initial $\chi^2_{\nu}$ value with a model including no magnetic 
field is of the order of 2. However,
our observations were collected at four epochs only, implying 
that the reconstructed image, as well as the conclusion concerning the
rotation period of 3.1d, must be taken with caution, and need to be confirmed
with further observations.

We obtain a field of only a few Gauss at the star's surface, which
is one of the weakest stellar magnetic fields detected so far.
The solution presented in Figs.~\ref{map} and~\ref{fit} includes a 
toroidal component of the magnetic field. Introducing such a component 
leads to a field configuration containing about 15\% less information
than when no toroidal field is assumed,
indicating that it is likely present, although more data would be 
necessary to confirm its 
existence. The fit of the Stokes $V$ profiles near orbital phase 1.7 is not
entirely satisfactory, indicating that the magnetic field is probably 
slightly more complex than our model. Pushing the reconstruction procedure 
further to improve the fit near this phase would probably be poorly 
significant, since the reduced $\chi^2_{\nu}$ is already near 1.
Only more complete data with better phase coverage will 
help improve the magnetic image.

When the 18\% relative differential rotation discussed in Sect. 3.2 is 
introduced in the
reconstruction, we find a minimum of $\chi^2_{\nu}$ for an equatorial 
period of 3.0d. The reconstructed image is identical to the one with no
differential rotation, the configuration implying a rotation period of 
3.1d at the mean latitude of the magnetic spots. In this configuration, the
rotation period at the pole is 3.7d. We noted that the reconstructed image
assuming an 18\% differential rotation contains 10\% less information 
than images assuming no differential rotation or a 30\% differential rotation.
This result provides an independant
confirmation of the level
of differential rotation derived from line profiles in Sect. 3.2.

\section{Discussion and conclusion}

Our data clearly indicate a surface relative differential rotation 
$d\Omega/\Omega$ of about 18\%, which corresponds to 
$d\Omega \approx$ 0.3--0.4 rad.d$^{-1}$. 
This level of differential rotation, which is comparable to that observed 
in many other cool stars (e.g. Reiners 2006), is much higher than that of 
the Sun, and also significantly higher than predicted by 
K\"uker \& R\"udiger (2005)
for F-type stars. This behaviour is also discussed by Marsden et al. (2006)
and Jeffers et al. (2006).
We find that the Stokes $V$ profile variations are best
modelled with a modulation period of 3.1d and an inclination angle of 
about 40$^{\circ}$. In this case, the differential rotation of $\tau$~Boo 
implies an equatorial (resp. polar) rotation period of 3.0d (resp. 3.7d). 
It is interesting to note that this
range of periods between pole and equator is not very different from that 
found by Henry et al. (2000) from spectrophotometric monitoring of the 
Ca~{\sc II} H \& K lines over several years, possibly interpreted as 
an expression of surface differential rotation.
The modulation period of 3.1d corresponds to the star's rotation at 
a latitude of about $25^{\circ}$. 


We have detected a weak magnetic field at the surface of $\tau$~Boo, with
intensities of only a few Gauss, i.e. similar to that of the Sun if it was
observed in the same conditions. On the other hand, the magnetic field 
topology of $\tau$~Boo, even in the simplified description derived from our 
limited data set, seems more complex than that of the Sun.
The reconstructed magnetic image of $\tau$~Boo indicates a dominant poloidal
field, with the probable presence of a small toroidal component. More data 
would be needed to confirm the existence of this toroidal component.
The differential rotation of $\tau$~Boo stengthens the idea that this component
is present, since field toroidal components are often associated with surface 
differential rotation (Donati et al. 2003, 2006).
The modulation period of the magnetic signature ($3.1 \pm 0.1$d) 
seems to be different from the planet orbital period (3.31d). 
However, our limited data do not provide strong 
constraints on the magnetic topology, and in particular are insufficient
to allow us to model the whole magnetosphere, using field extrapolation 
techniques (e.g., Jardine et al 2002), and study how the giant planet 
may interact with the stellar magnetosphere and possibly trigger activity 
enhancements correlated with the planet orbital motion.



The details of the star's magnetosphere and its potential interaction 
with the planet clearly need to be studied in the future. We note that 
a similar study for other planet-hosting stars would be of major interest,
in particular for those stars for which activity in the 
Ca~{\sc II} H \& K lines is observed, and possibly correlated with 
the planet orbital motion, such as HD~179949 or $\upsilon$~And (Shkolnik et al.
2003, 2005). Such studies will require data 
providing a much better phase coverage and recorded on a longer time 
scale than those presented here. Long spectropolarimetric monitoring of 
$\tau$~Boo and other planet-hosting stars will also be necessary to measure 
precisely their rotation and differential rotation. 

\section*{Acknowledgments}

We warmly thank the CFHT staff for their efficient help during the 
observations. We are indebted to an anonymous referee for very valuable 
comments that led to a significant improvement of this paper.

\end{document}